# What does ethnic density represent? Spatial co-occurrence networks of a widely used contextual measure using harmonised UK small-area census data

**Dr Joseph Lam**[a]

a. Great Ormond Street Institute of Child Health, University College London, 30 Guilford St, London WC1N 1EH

joseph.lam@ucl.ac.uk

**Acknowledgements**

This programme is supported in part through UKRI's Securing Better Health, Ageing and Wellbeing strategic theme. This work was supported by Health Data Research UK (HDRUK2023.0029, HDRUK2025.0427), which is funded by the Medical Research Council (UKRI), the National Institute for Health Research, the British Heart Foundation, Cancer Research UK, the Economic and Social Research Council (UKRI), the Engineering and Physical Sciences, Research Council (UKRI), Health and Care Research Wales, Chief Scientist Office of the Scottish Government, Health and Social Care Directorates, and Health and Social Care Research and Development Division (Public Health Agency, Northern Ireland). This work is (partly) funded by the NIHR GOSH BRC. The views expressed are those of the author(s) and not necessarily those of the NHS, the NIHR or the Department of Health.

What does ethnic density represent? Spatial co-occurrence networks of a widely used contextual measure using harmonised UK small-area census data

## Abstract

**Background**: Ethnic density is widely used in epidemiology and health geography as a contextual exposure, usually defined as the proportion of local residents who share an individual's ethnic group. Yet its contextual interpretation is rarely examined directly. The contextual correlates of equivalent percentage values may differ across groups and places, particularly where migration, religion, language, household structure and socioeconomic conditions are spatially co-located.

**Methods**: Using the harmonised Unified UK Census Data release, I analysed 239,023 small areas to examine the contextual co-occurrence surrounding ethnic density, defined as area-level covariation between ethnic composition and selected demographic, migration, language, religion, household, housing and socioeconomic characteristics. I estimated separate UK-wide mixed graphical models (MGM) for eight ethnic-density targets, using 239,019 complete cases and 32 nodes in each target-specific model. England-only spatial analyses used k-nearest-neighbour Output Area centroids (k = 8) to estimate Global Moran's I, Local Moran's I and sensitivity networks after regional and local spatial adjustments.

**Results**: Ethnic density did not behave as a single contextual scalar. The strongest conditional ecological associations differed across the group-specific ethnic density measures. In the UK-wide MGM, Asian density was associated most strongly with Middle East/Asia-born share (0.59), Indian density with Hindu share (0.55), Pakistani density with Muslim share (0.47), Bangladeshi density with Muslim share (0.23), Black density with Africa-born share (0.42), Other density with Muslim share (0.23), Mixed density with multi-ethnic households (0.25), and White density with Middle East/Asia-born share (0.35). England-only ethnic-density measures were strongly spatially autocorrelated, with Global Moran's I ranging from 0.57 for Mixed share to 0.90 for White share. After residualising against English region and local spatial lag, 64.3% to 96.4% of original edges were retained in the spatially adjusted sensitivity networks.

**Conclusion**: Ethnic density is a compositional measure that can be embedded within different social and spatial neighbourhood contexts across groups and places. Equivalent percentages therefore need not describe equivalent wider neighbourhood contexts, although the relevance of these differences depends on the research question and hypothesised mechanism. The UK-wide and England-wide analyses are not an urban-only study; they illustrate a network and spatial workflow with potential application in urban health research.

## Introduction

Ethnic density is one of the most recognisable contextual variables in research on ethnicity and health (1). In most empirical applications it is operationalised as the proportion of residents in a local area who belong to the same ethnic group as the study participant, that is, own-group or co-ethnic density (2,3). That measure is then entered into individual-level or multilevel regression models as a neighbourhood exposure, often alongside the person's own ethnicity, socioeconomic characteristics, and one or more area-level covariates. In the UK, ethnic density has been especially influential in work on adult mental health and psychosis, where living among more co-ethnics can be protective across multiple systematic reviews and meta-analyses (4–8). The evidence is less consistent for other health behaviours and outcomes (9–13). Across these studies, ethnic density is generally treated as a proxy for one or more neighbourhood mechanisms: reduced interpersonal racism and minority stress; greater social support and collective efficacy; stronger institutional completeness, including religious and cultural infrastructure; or differences in settlement history, migration recency and language ecology.

Yet this literature contains a persistent conceptual tension. Ethnic density is treated as if it were a stable neighbourhood exposure, but the same numerical value can plausibly represent very different local contextual, social worlds (2). Many ethnic density studies adjust for area deprivation and other individual- or area-level characteristics, yet the mechanisms through which ethnic density operates remain poorly understood (10,13). This means that adjustment alone does not resolve the prior measurement question of what ethnic density represents at neighbourhood scale, and raises the possibility that some routinely adjusted variables may overlap with the contextual bundle the measure is intended to capture. Here, I use contextual co-occurrence to refer to the ecological co-location, within the same small-area geography, of ethnic composition with other neighbourhood characteristics. This problem matters for both geography and epidemiology because ethnic density is not simply a compositional statistic. From a geographical perspective, ethnic density reflects spatially organised neighbourhood processes that differ substantially in migration history, religion, language, household structure, socioeconomic conditions and wider settlement geography (14). From an epidemiological perspective, whether those differences are relevant depends on the research question and hypothesised mechanism. If ethnic density is being used principally as a measure of potential exposure to co-ethnic or non-co-ethnic others, some accompanying neighbourhood characteristics may be peripheral to the exposure of interest (15). If it is intended as a proxy for a broader social environment involving community support, institutional resources or settlement context, those same characteristics may be integral to its interpretation (4). The prior measurement question is therefore not whether ethnic density has one universally correct meaning, but what wider neighbourhood characteristics accompany a given level of ethnic density, and whether those patterns are similar across groups and places.

The arrival of a new harmonised small-area UK census dataset makes it possible to address this measurement gap more directly (16). Goodwin and Singleton's unified dataset harmonises 2021 England, Wales and Northern Ireland census tables with the 2022 Scotland census (17). The release contains harmonised small-area counts for ethnicity, country of birth, year of arrival, English-language proficiency, religion, housing, household structure, and several socioeconomic domains. This creates an opportunity to ask a prior measurement question that most ethnic density studies leave implicit: what neighbourhood characteristics co-locate with ethnic density?

This framing also connects ethnic density to a wider measurement problem in neighbourhood and urban health research: contextual exposures are rarely encountered as isolated environmental or social conditions (14,18). The same percentage of green space may refer to public parks, private gardens, inaccessible land, tree canopy, sports fields, or unmanaged open space, each with different health relevance (19). Similarly, the Index of Multiple Deprivation is explicitly multidimensional, yet different combinations of its domains were often collapsed into a singular representation of "deprivation". These examples illustrate a wider measurement issue: equivalent values of a neighbourhood variable may not be substantively equivalent. Ethnic density is a useful empirical case because it can be expressed as a single percentage, yet its interpretation may depend on the other contextual features with which that percentage co-occurs. Examining these underlying configurations before outcome modelling helps clarify what kind of neighbourhood construct is being measured, whether apparently equivalent exposure levels are comparable, and how findings should be interpreted in relation to urban health inequalities.

This paper therefore treats ethnic density as a construct to be examined before it is inserted into an outcome model. The main argument is that ethnic density is a simple compositional measure, but the neighbourhood context accompanying a given value need not be simple or equivalent across ethnic groups and places. Using harmonised small-area census data, it asks four questions. First, which contextual characteristics co-occur with different ethnic density measures? Second, do those co-occurrence profiles differ across ethnic groups when analysed as conditional dependency networks? Third, how are the ethnic density measures spatially clustered at local scale? Fourth, do conditional associations with ethnic density remain after accounting for broad English region and local spatial dependence? Together, these analyses use ethnic density to illustrate a broader diagnostic approach to contextual measurement: network analysis can describe which neighbourhood characteristics remain associated with a focal measure conditional on the other measured characteristics, while spatial analysis can assess how strongly those patterns are organised geographically. The purpose is not to establish causal pathways or separability, but to make the wider contextual embedding of an apparently singular area-level measure visible before it is used in an outcome model.

## Data and methods

### Study design

The study is an ecological analysis of neighbourhood composition and contextual co-occurrence. Contextual co-occurrence refers to area-level covariation between ethnic composition and other social, demographic, housing and socioeconomic characteristics measured within the same small-area geography; it does not imply that these characteristics co-occur within the same individuals or households. It examines which wider neighbourhood characteristics are associated with different ethnic density measures across harmonised small areas, and how these associations are spatially patterned.

The analysis proceeded in four stages. First, I constructed an ecological correlation matrix across all ethnic density and contextual variables to describe which neighbourhood characteristics tended to co-vary across small areas. Second, I estimated separate conditional dependency networks for each ethnic density measure using mixed graphical models. Each network contained one focal ethnic density variable together with the same set of contextual variables, allowing the conditional associations surrounding each ethnic density measure to be compared across groups. Third, I assessed the spatial clustering of the ethnic density measure in England using Local Indicators of Spatial Association (LISA). Fourth, I re-estimated the conditional dependency networks after removing broad regional variation and local spatial patterning from each variable, to assess which associations between ethnic density and contextual characteristics remained.

The harmonised UK dataset supports the primary UK-wide correlation and network analyses. The LISA and spatial analyses were restricted to England OAs, because these analyses required a consistent small-area geography, English region identifier, and consistent OA 2021 centroid coordinates. This design separates the primary UK-wide analysis of contextual co-occurrence from an England-only analysis of its spatial patterning and sensitivity to regional and local spatial structure.

### Data source and harmonisation context

The data source was the unified small-area UK census release (18). The dataset harmonises tables released separately by the Office for National Statistics, National Records of Scotland, and the Northern Ireland Statistics and Research Agency. The harmonised data contain 190 unified variables organised into 25 topic tables and indexed across 239,023 harmonised small areas. The release is limited to univariate small-area tables because multivariate tables are not consistently available at the same fine spatial scale across the UK. The dataset is well suited to ecological co-location analysis, but it cannot identify whether characteristics observed within the same area co-occur within particular individuals or households.

Goodwin and Singleton also emphasised two technical issues arising from census harmonisation that shaped the analysis. First, table totals should not be treated as a single definitive OA population count. Disclosure control and perturbation were applied differently across the source agencies, meaning household and population totals can differ across tables even when they appear to refer to the same conceptual population. Second, only a minority of matched tables were directly equivalent; most required aggregation for comparability. Accordingly, all derived proportions in this study use table-specific denominators, and interpretations are restricted to harmonised ecological measures rather than exact equivalents of the original nation-specific classifications.

## Geography and spatial weights

For England, the spatial analyses use OAs, which typically contain 40-250 households and 100-625 residents (19). I used the OA 2021 to region lookup for England and official ONS OA 2021 population-weighted centroids in British National Grid (EPSG:27700). Local spatial relationships between OAs were represented using a k-nearest neighbours graph based on official ONS population-weighted centroids.

The primary neighbourhood definition was k-nearest neighbours with k = 8 on the centroid coordinates, providing each OA with a small local comparison set while avoiding zero-neighbour areas. Because k-nearest-neighbour relationships are not necessarily reciprocal, the neighbour graph was symmetrised: two OAs were treated as neighbours if either OA identified the other among its eight nearest neighbours. After symmetrisation, OAs had a median of 9 neighbours (range 8–21). This provides a complete and reproducible local spatial structure for all 178,605 England OAs, but it should be interpreted as a centroid-distance approximation to local spatial adjacency rather than as a boundary-contiguity definition.

## Measures

### Ethnic Density

Ethnic density measures were derived from the harmonised ethnic group table (uk021) using the table total as denominator (17). Eight targets were examined: White; Asian; Indian; Pakistani; Bangladeshi; Black, Caribbean or African (Black); Mixed or multiple ethnic groups (Mixed), and Other ethnic group (Other) share. These targets are not mutually exclusive analytical categories: the broad Asian measure overlaps conceptually with the Indian, Pakistani and Bangladeshi subgroup measures. For England, Wales and Scotland, the harmonised Asian measure maps to the corresponding published broad Asian total; for Northern Ireland, it is constructed by aggregating the available Indian, Chinese, Filipino, Pakistani and Other Asian categories. The broad and subgroup measures were therefore analysed in separate target-specific models rather than entered together. The harmonised Bangladeshi subgroup has no directly matched Northern Ireland source category, reflecting differences in the ethnic classifications available across the four national censuses.

These groups were selected for three reasons. First, they reflect policy-relevant and substantively recognisable categories in UK ethnic inequalities research, including the ethnic density literature in psychiatry and population health, which has typically focused on White-majority context alongside Black, South Asian, and, where data permit, more specific subgroup densities (8,9). Previous UK studies have examined Indian, Pakistani and Bangladeshi own-group density separately, reflecting the possibility that ethnic-density associations and their underlying social contexts differ between these populations (7,20). Second, they correspond to categories that remain sufficiently robust after UK-wide harmonisation of the four national censuses (17). This is particularly important because the unified small-area dataset necessarily aggregates some categories differently across nations, making fine-grained subgroup comparisons less defensible. Third, these groups provide a balance between interpretability and statistical stability: they are large enough to support meaningful small-area analysis across the UK, while still allowing comparison between broader pan-ethnic categories and selected more specific groups, such as Indian, Pakistani and Bangladeshi, whose neighbourhood patterning may differ substantially. Census ethnic categories should not be interpreted as fixed or homogeneous identities; previous longitudinal research has demonstrated both variation in the stability of individual ethnic-group responses and a trade-off between greater stability at broader levels of aggregation and loss of substantive detail (21,22).

**Exploratory contextual variable set**

The contextual variables were selected at the level of substantive domains rather than as a collection of hypothesised causal covariates (Table 1). Neighbourhood-effects research conceptualises place as comprising interrelated compositional, material and social characteristics rather than a single contextual dimension (14,19). Ethnic-density studies similarly implicate migration and settlement, social and cultural context, and socioeconomic and material conditions as potential mechanisms, correlates or adjustment factors, while evidence on the mechanisms underlying observed ethnic density associations remains incomplete (6,7,10). Research on UK ethnic residential geographies further demonstrates that ethnic composition is spatially patterned alongside settlement histories and changing household and neighbourhood structures (23,24). These literatures informed six broad domains available within the harmonised census resource: migration and settlement history; language; religion; household and demographic structure; housing and material conditions; and socioeconomic position. The aim was not to assign these characteristics a priori as confounders, mediators, or components of ethnic density, but to define a broad contextual field within which their ecological co-occurrence with ethnic composition could be examined.

| Domain | Derived measures | Rationale for inclusion |
|---|---|---|
| Migration and settlement history | UK-born; Africa-born; Middle East/Asia-born; arrived since 2001; arrived since 2011 | Migration generation, nativity and settlement history are repeatedly considered in ethnic density research and may shape the social meaning of co-ethnic concentration, including through residential settlement, social networks and acculturation (6). |
| Language | Limited English proficiency | Language is related to migration and settlement processes and may influence opportunities for social interaction, access to services and the degree to which co-ethnic concentration corresponds to a shared linguistic environment (10). |
| Religion | No religion; Christian; Hindu; Muslim; other religion | Religious composition is relevant to the cultural and social organisation of local populations and may capture an aspect of the community context accompanying ethnic concentration. Religious affiliation is treated here as an area characteristic rather than a direct measure of religious institutions or institutional completeness (6,25). |
| Household and demographic structure | Children; young adults; older adults; one-person households; lone-parent households; large households (5+); multi-ethnic households | Household composition, age structure and ethnic mixing vary geographically and are related to migration, family formation and residential settlement. Including these measures allows ethnic density to be situated within broader demographic and household configurations rather than treated as an isolated population share (23,24) |
| Housing and material conditions | Flats or tenements; social rent; private rent; owner occupation; no-car households | Housing and material conditions are established dimensions of neighbourhood context and are closely linked to residential opportunity, socioeconomic position and the spatial concentration of population groups (14,19). |
| Socioeconomic position | High and low National Statistics Socio-economic Classification (NS-SEC); unemployment; inactivity; professional occupations; elementary occupations; no qualifications; degree-level qualifications | Socioeconomic conditions and area deprivation are among the most commonly considered contextual factors in ethnic-density studies and may both pattern residential composition and modify or account for observed ethnic density associations (6,7) |

**Table 1. Selected domains of contextual co-occurrence with operationalised variables in harmonised census.**

All measures were expressed as area-level proportions using the denominator appropriate to their source census table. For ethnic group, country of birth, religion and age variables, denominators were usual residents or the relevant age-specific resident population. For household composition, tenure, accommodation, car access and multi-ethnic-household variables, denominators were households or household spaces. For economic activity, occupation and qualifications, denominators were the relevant eligible population, such as residents aged 16 years and over in employment. For readability, these measures are described as shares, but each uses its source-table-specific denominator.

## Analytical strategy

### Ecological correlation

First, I described ecological co-occurrence between ethnic density and the broad harmonised contextual variable set using pairwise correlation across small areas. These summaries describe which neighbourhood characteristics tend to vary together with each ethnic density measure across small areas.

### Conditional dependency networks

Second, I estimated separate conditional dependency networks for each ethnic density measure using mixed graphical models (MGM). A graphical model represents variables as nodes and conditional associations between pairs of variables as edges. Each UK-wide network contained one focal ethnic density measure with the same 31 contextual measures, giving 32 nodes in total. The broad and subgroup ethnic-density measures were therefore analysed in separate models rather than simultaneously. An edge between ethnic density and a contextual characteristic indicates that the two remained associated after accounting for the other variables included in that network. English region was not included in the UK-wide networks because an equivalent regional classification was not available consistently across all four UK nations.

The UK-wide analysis used 239,019 complete small areas. Proportion variables were transformed using an empirical logit where appropriate and modelled as continuous Gaussian variables to reduce problems arising from bounded or skewed distributions. MGM estimates a regularised network in which weaker conditional associations can be shrunk to zero. The degree of regularisation was selected using the extended Bayesian information criterion (EBIC) with $\gamma = 0.25$, rather than by applying an arbitrary post-estimation threshold to edge weights. Non-zero pairwise associations retained by the fitted model were represented as network edges. Models included pairwise interactions only ($k = 2$) and were estimated with the mgm package (26).

**Spatial clustering: Global Moran’s I and Local Moran’s I**

Third, I assessed spatial clustering of each ethnic density measure in England using Global Moran's I and Local Moran's I. Global Moran's I measures overall spatial autocorrelation; positive values indicate that OAs with similar values tend to be located near one another. Local Moran's I identifies where this pattern is concentrated by classifying significant local clustering. OAs were classified as High-High, Low-Low, High-Low and Low-High clusters. High-High clusters have relatively high ethnic density values surrounded by high-value neighbours; Low-Low clusters have relatively low ethnic density values surrounded by low-value neighbours; High-Low and Low-High clusters indicate local spatial outliers. LISA classifications were interpreted descriptively as indicators of local spatial patterning.

**Spatially adjusted residual networks**

Fourth, I used spatially adjusted residual networks as a sensitivity analysis to examine whether the conditional associations identified in England remained after accounting for broad regional differences and local spatial patterning. For a like-for-like comparison, I estimated unadjusted England-only MGM networks using the same variable set before residualisation. Edge persistence was assessed by comparing these England-only networks with the corresponding spatially adjusted residual networks. For each contextual and ethnic density variable, I fitted a separate model including English region and the variable’s spatial lag, defined as the weighted mean of values in its *k*-nearest-neighbour OAs. The residual from this model represents variation not accounted for by broad regional location or local spatial similarity. Residuals were standardised and entered into the same MGM procedure used for the primary networks. An ethnic density contextual edge that remained in the residual network was interpreted as a conditional ecological association that was not wholly attributable to these broad regional and local spatial patterns. Conversely, disappearance of an edge does not indicate that the original association was spurious or disproven; it indicates that the association was sensitive to the spatial structure removed by the residualisation procedure.

**Software and reproducibility**

Data processing, modelling and visualisation were conducted in R 4.6.0 (27). Spatial neighbourhood objects, Global Moran's I and Local Moran's I were implemented using spdep (28); sf was used for spatial data handling where needed. Figures were produced with ggplot2 and qgraph (29,30). Code for replication is available on the author’s GitHub Repository (31).

## Results

The ecological correlation summaries showed distinct descriptive profiles for each ethnic density measure before conditional network modelling (Figure 1). Asian, Pakistani and Bangladeshi density were each positively correlated with Muslim share, Middle East/Asia-born share, limited English and large-household share, although the magnitude of these correlations differed across groups. Asian density was most strongly correlated with Middle East/Asia-born share (r = 0.90) and Muslim share (r = 0.84), followed by limited English (r = 0.63) and large households (r = 0.62). Pakistani density was most strongly correlated with Muslim share (r = 0.81), large households (r = 0.56) and Middle East/Asia-born share (r = 0.55), whereas Bangladeshi density was most strongly correlated with Muslim share (r = 0.55), Middle East/Asia-born share (r = 0.40), large households (r = 0.36) and limited English (r = 0.36). Indian density was most strongly correlated with Hindu share (r = 0.80) and Middle East/Asia-born share (r = 0.68), while Black density was most strongly correlated with Africa-born share (r = 0.84) and arrival since 2001 (r = 0.53). Mixed density was most strongly correlated with multi-ethnic-household share (r = 0.58). Other density was negatively correlated most strongly with UK-born share (r = -0.70) and White share (r = -0.69). White density showed the opposite compositional pattern, with positive correlation with UK-born share (r = 0.84) and negative correlations with Middle East/Asia-born share (r = -0.85), Muslim share (r = -0.82), arrival since 2001 (r = -0.74) and Africa-born share (r = -0.71). These descriptive correlations provide an initial account of ecological co-location across small areas and the descriptive context for the conditional network analysis.

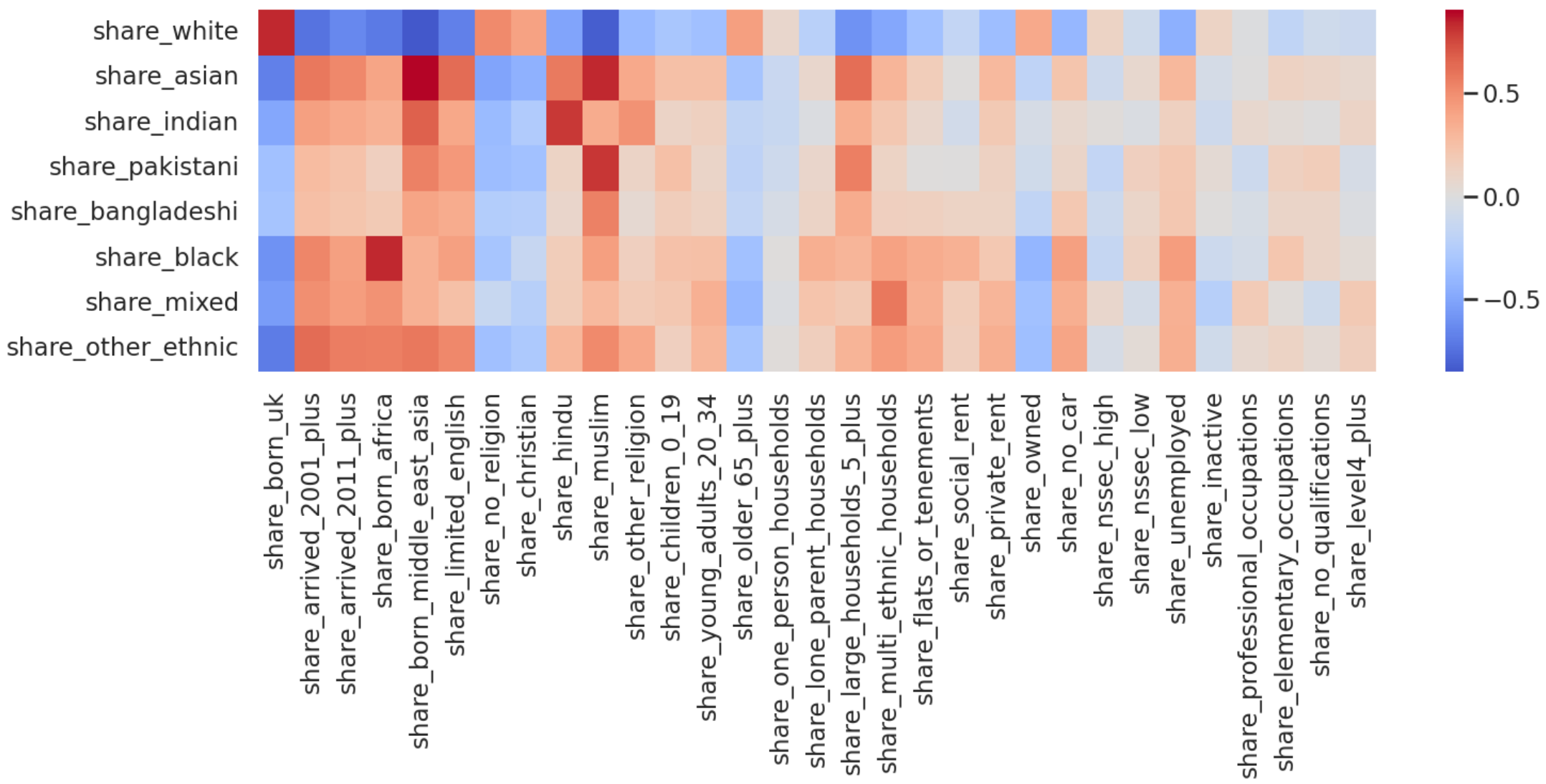

**Figure 1. Correlation between ethnic density and neighbourhood features at small-area level.**

### Contextual co-occurrence and conditional network structure

The UK-wide network analysis showed that the conditional associations surrounding each ethnic density measure differed across groups (Figure 2). For each ethnic density model, we report the five largest absolute non-zero conditional associations with the focal ethnic density node, providing a concise comparison of the most prominent contextual characteristics associated with each measure. Full edge tables are reported in Supplementary Table S1.

Asian density was most strongly conditionally associated with Middle East/Asia-born share, followed by Hindu and Muslim share. Indian density was most strongly associated with Hindu share and Middle East/Asia-born share. Pakistani and Bangladeshi density were both most strongly associated with Muslim share, although their other leading conditional associations differed in magnitude and composition. Black density was most strongly associated with Africa-born share, while Mixed density was most strongly associated with multi-ethnic households and higher socioeconomic-position measures. Other ethnic share was most strongly associated with Muslim share, followed by other religion share and multi-ethnic households. White density had leading conditional associations with several markers of population composition, including Middle East/Asia-born share, Muslim share, no-religion share, Africa-born share and multi-ethnic households.

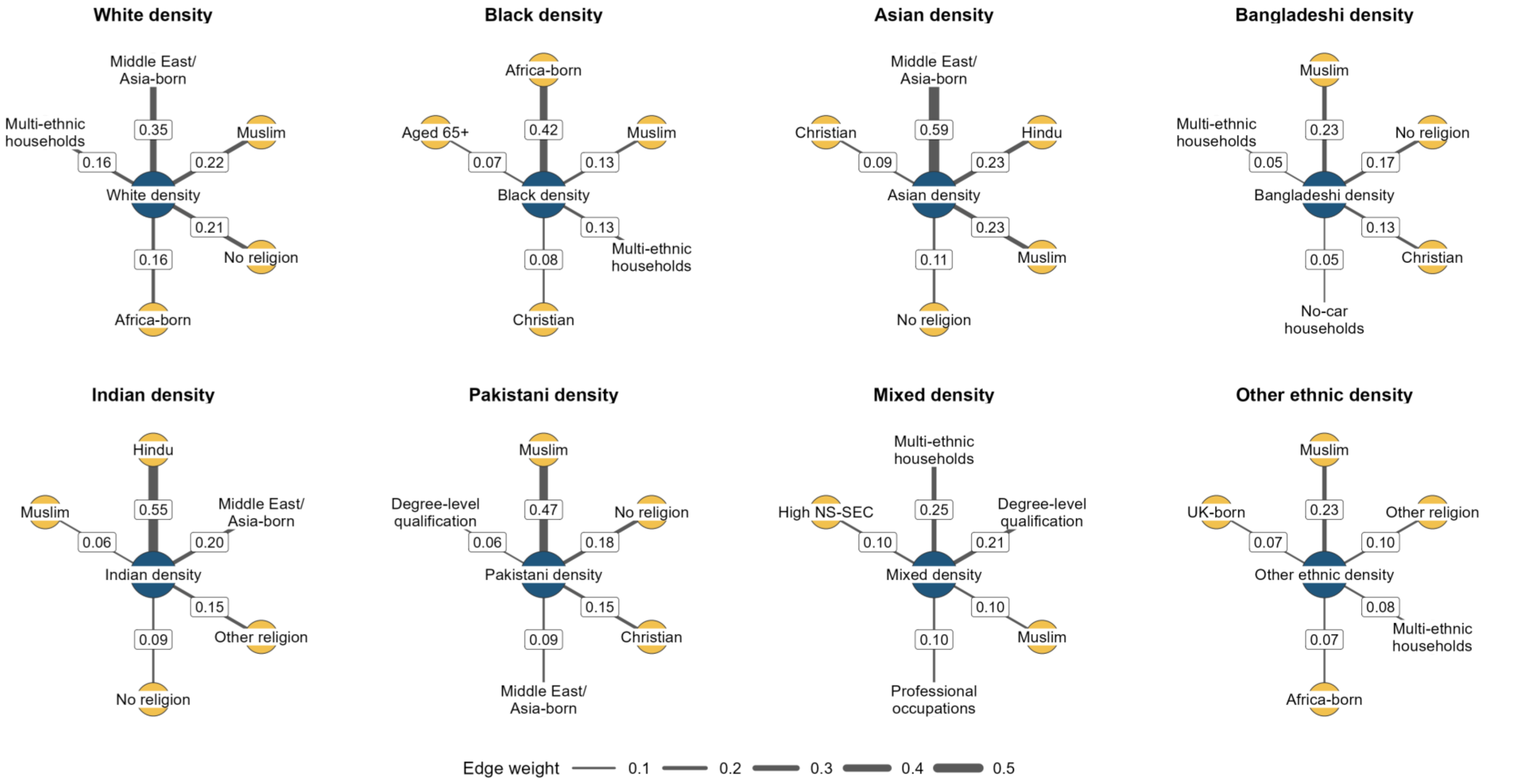

**Figure 2. Five strongest conditional associations with each ethnic density measure in UK-wide mixed graphical models of contextual co-occurrence. Each panel shows one focal ethnic density node and the five strongest absolute non-zero edge weights from the corresponding mixed graphical model. Edge labels show regularised mixed graphical model edge weights, with thicker lines indicating larger absolute edge weights.**

## Spatial clustering of ethnic density measures

The spatial diagnostics showed that ethnic density measures were strongly spatially patterned (Table 2). White density had the strongest global spatial autocorrelation (Moran's I = 0.90), followed by Asian (0.86), Pakistani (0.83), Indian (0.81) and Bangladeshi (0.75) density. Local Moran's I showed that the form of spatial patterning differed across groups. White density had the largest proportion of areas classified as significant LISA clusters or spatial outliers (49.1%), including both High-High and Low-Low clusters, indicating broad spatial differentiation between areas with high-White-share and lower-White-share. In contrast, Indian, Pakistani, Bangladeshi and Black density showed strong global spatial autocorrelation but smaller proportions of significant LISA areas, largely concentrated in High-High clusters. The LISA maps (Figure 3) showed broad local clustering for White and Mixed density, but more geographically concentrated High-High clusters for Black, Pakistani and several other minority ethnic density measures. Mixed density showed a different pattern, with lower global autocorrelation but a comparatively high proportion of significant local clusters, including both High-High and Low-Low areas.

| Ethnic density target | Global Moran's I | Significant LISA (%) | High-High (%) | Low-Low (%) |
|---|---|---|---|---|
| White | 0.900 | 49.1 | 29.2 | 19.6 |
| Asian | 0.862 | 16.5 | 13.0 | 3.0 |
| Indian | 0.813 | 9.2 | 8.7 | 0.0 |
| Pakistani | 0.833 | 7.6 | 7.2 | 0.0 |
| Bangladeshi | 0.746 | 4.9 | 4.4 | 0.0 |
| Black | 0.772 | 14.1 | 13.0 | 0.3 |
| Mixed | 0.571 | 42.0 | 17.6 | 21.1 |
| Other ethnic | 0.683 | 18.3 | 14.1 | 3.0 |

**Table 2. Spatial clustering summary for ethnic-density measures.**

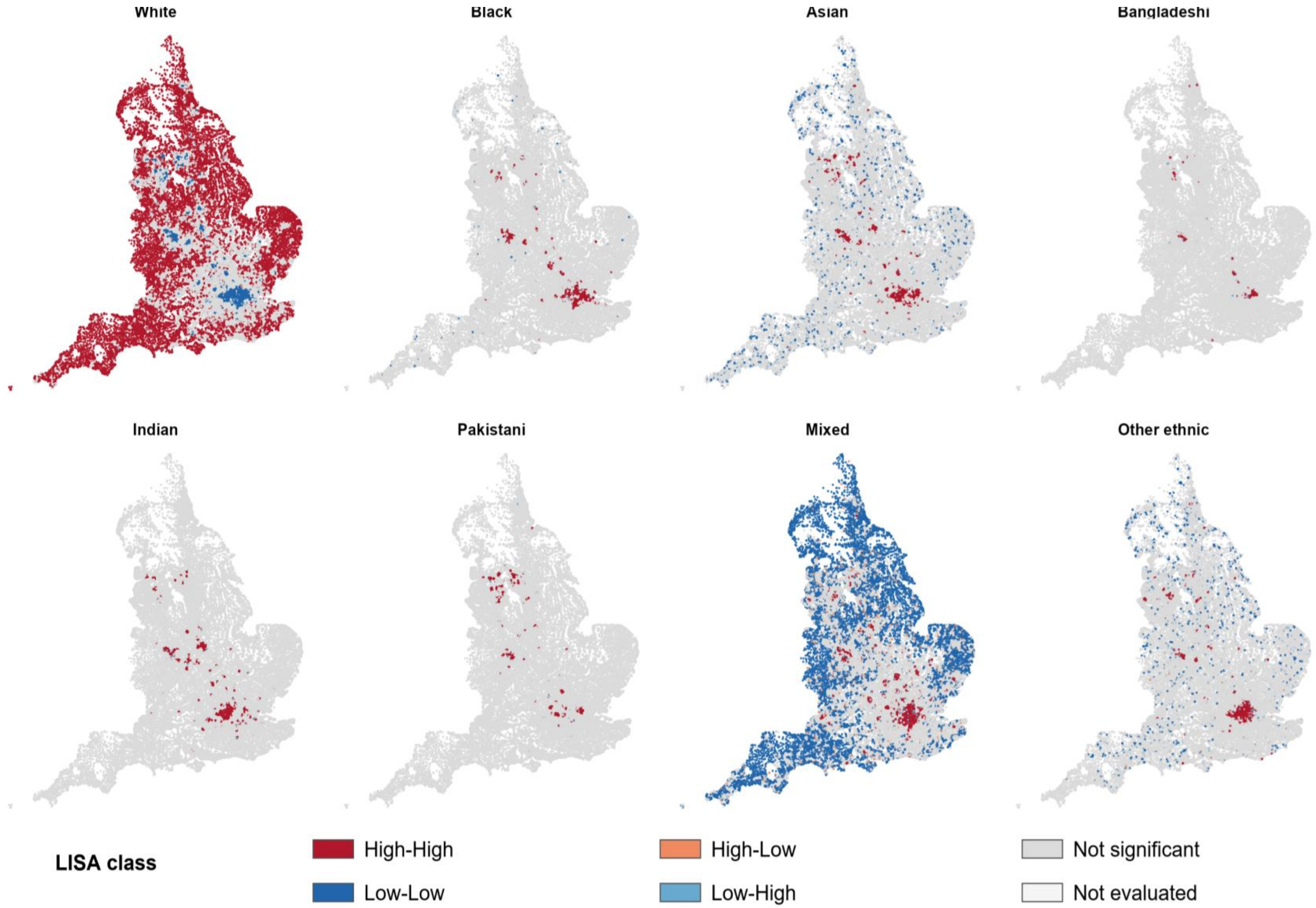

**Figure 3. Local Moran's I cluster maps for ethnic-density measures. High-High clusters indicate OAs with high values surrounded by high-value neighbours; Low-Low clusters indicate spatial concentration of low values; High-Low and Low-High areas indicate local spatial outliers.**

## Spatially adjusted residual networks

The spatially adjusted residual networks provided a sensitivity analysis of the conditional network findings. Most leading conditional associations with the ethnic density nodes attenuated after accounting for English region and local spatial lag, although several of the strongest group-specific associations persisted (Figure 4; Supplementary Table S2 for top five persistent edges). These included Asian density with Middle East/Asia-born share, Indian density with Hindu share, Pakistani and Bangladeshi density with Muslim share, Black density with Africa-born share, and Mixed density with multi-ethnic-household share. White density also retained leading associations with Middle East/Asia-born share and multi-ethnic-household share, while its religion-related associations were attenuated after spatial adjustment.

Edge persistence varied across ethnic-density networks (Table 3). Black density showed the highest persistence, with 27 of 28 original edges involving the ethnic density node retained after spatial adjustment (96.4%), followed by Mixed density (25 of 27; 92.6%) and White density (25 of 28; 89.3%). Persistence was lower for Asian density (18 of 28; 64.3%) and Other ethnic density (18 of 25; 72.0%), indicating that a larger proportion of their original conditional associations was sensitive to regional and local spatial patterning.

Edges that did not persist after spatial adjustment were predominantly smaller associations involving socioeconomic position, housing, car access and household structure. For example, Asian density no longer showed conditional associations with degree-level qualification, private rent, one-person households, social rent and unemployment. Pakistani density no longer showed conditional associations with no qualifications, flats or tenements, no-car households, social rent, children aged 0-19 and limited English. Other ethnic density no longer showed several smaller socioeconomic and housing-related edges, including degree-level qualification, high NS-SEC, inactivity, no-car households, social rent and private rent. By contrast, relatively few associations were lost from the Black and Mixed density networks: Black density lost only its association with degree-level qualification, while Mixed density lost associations with no-car households and elementary occupations.

| Ethnic density group | Unadjusted England-only edges | Adjusted edges | Persisted | Dropped | Persistence (%) |
|---|---|---|---|---|---|
| White | 28 | 26 | 25 | 3 | 89.3 |
| Asian | 28 | 19 | 18 | 10 | 64.3 |
| Indian | 24 | 21 | 19 | 5 | 79.2 |
| Pakistani | 29 | 23 | 22 | 7 | 75.9 |
| Bangladeshi | 28 | 24 | 22 | 6 | 78.6 |
| Black | 28 | 30 | 27 | 1 | 96.4 |
| Mixed | 27 | 28 | 25 | 2 | 92.6 |
| Other ethnic | 25 | 21 | 18 | 7 | 72.0 |

**Table 3. Persistence of England-only edges after spatial adjustment.**

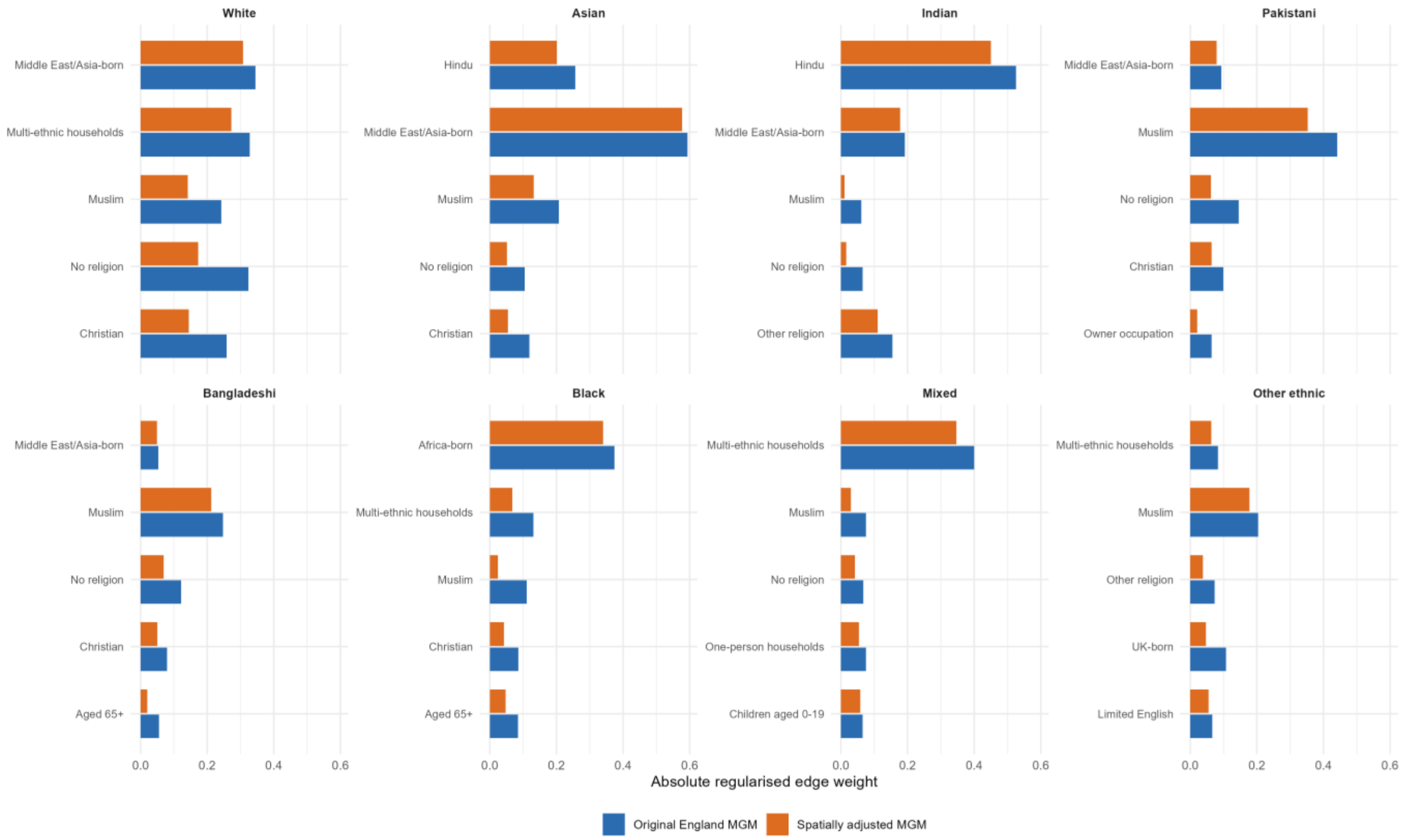

**Figure 4. Five strongest conditional associations with each ethnic density measure in England before and after spatial adjustment.**

## Discussion

This study suggests that ethnic density, although conventionally expressed as a single area-level proportion, does not have a stable contextual meaning across ethnic groups or places. The neighbourhood characteristics accompanying higher ethnic density differed across groups, spanning migration, religion, language, household structure, socioeconomic context and wider settlement patterns. The UK-wide network analysis showed that the strongest conditional associations surrounding ethnic density varied substantially across groups: Asian, Indian, Pakistani, Bangladeshi, Black, Mixed, Other and White density were each associated with different configurations of migration, religion, household and socioeconomic characteristics. The England spatial analyses further showed that ethnic density measures differed in their degree and form of spatial clustering, while many of the leading conditional associations remained after adjustment for English region and local spatial lag. These findings indicate that the same area-level measure, proportion of residents belonging to a particular ethnic group, can be embedded within materially different neighbourhood contexts depending on the group and place under study.

These results do not undermine the published ethnic density literature. Rather, they may help clarify what the associations reported in the literature represent. For example, previous studies demonstrated that higher own-group density was associated with lower risk of common mental disorder for some minority ethnic groups, although reduced discrimination and greater social support did not fully account for these associations, and the protective associations were not observed universally across groups (8). The present findings add an ecological perspective to their interpretation by suggesting that higher ethnic density is accompanied by different neighbourhood characteristics for different groups. For one group, higher ethnic density may reduce social isolation in areas characterised by established migrant settlement; for another, it may coincide more strongly with religious composition, household structure or linguistic environments (32–34). Heterogeneity in associations reported across previous meta-analyses may therefore reflect not only differences in effect size, but also differences in the wider neighbourhood context accompanying the ethnic density measure.

A central implication is that ethnic density should not be assumed to have a single invariant interpretation. In much applied work, it is treated as a recognisable neighbourhood scalar: a percentage that can be entered into a regression model and interpreted as “more” or “less” co-ethnic context. Yet equivalent percentage values need not necessarily describe equivalent neighbourhood contexts across ethnic groups, because the characteristics accompanying those values can differ substantially (35). A 20% Pakistani area, a 20% Black area, and a 20% Mixed area are not merely different instances of the same wider social and spatial context. This ambiguity may be amplified when ethnic density is dichotomised into “high” and “low” categories, because the interpretation then also depends on the threshold used. Authors should therefore specify what role ethnic density is intended to play in their analysis: whether it represents overall exposure to a more co-ethnic local population, serves as a proxy for broader social environment, or is used principally as a descriptive measure of neighbourhood composition (22,36). The choice of accompanying covariates should follow from that interpretation and the hypothesised mechanism rather than from an assumption that there is a universally appropriate set of neighbourhood characteristics to adjust for.

The spatial analyses reinforce this interpretation by showing that ethnic density varies in both social content and spatial organisation. Studies that ignore this spatial structure may interpret ethnic density as a purely local neighbourhood characteristic when it may also reflect wider settlement geographies. The LISA maps illustrate this descriptively: High-High clusters for Black, Pakistani and Mixed density were concentrated in and around major urban areas, including London and parts of the West Midlands and Greater Manchester, while White-density Low-Low clusters were concentrated in more ethnically diverse urban areas, particularly around London. These patterns should not be interpreted as formal analyses of urban systems, or evidence that ethnic composition causes the observed spatial clustering. Rather, they show that ethnic density measures are embedded within broader geographical patterns that may extend beyond individual neighbourhood boundaries. Individual-level studies should therefore consider whether estimated ethnic density associations are sensitive to spatial scale, regional adjustment, urban concentration, and boundary definition rather than assuming that the relevant context is confined to immediate neighbourhood composition.

The broader methodological contribution of this paper is a diagnostic approach for characterising the social and spatial context surrounding an area-level measure. Conditional network analysis identifies which measured neighbourhood characteristics remain associated with the focal measure after accounting for the other variables in the network, while spatial analysis describes the extent and form of its geographical clustering. Neither analysis identifies causal pathways nor establishes that particular contextual dimensions are separable. Used together, however, they can make visible the wider contextual embedding of an apparently singular neighbourhood measure and identify features of that context that may be important for its substantive interpretation.

## Strengths and limitations

This study has several strengths. It uses a harmonised UK-wide small-area Census 2021/2022 resource to examine a measurement question that is rarely addressed directly in ethnic density research: which neighbourhood characteristics accompany variation in ethnic density before any health outcome is modelled. Rather than restricting the analysis to a single hypothesised mechanism, the study considers a broad set of social, demographic, housing and socioeconomic characteristics that have been implicated in previous ethnic density research. The use of mixed graphical models provides a further strength by identifying the conditional associations surrounding each ethnic-density measure after accounting for the other measured characteristics in the network.

Several limitations should also be noted. First, this is an ecological analysis of harmonised small-area counts and cannot establish individual-level mechanisms or causal pathways. The results describe which contextual characteristics co-vary with ethnic density across neighbourhoods, not whether those characteristics are experienced by the same individuals or households, or explain associations between ethnic density and health outcomes. Second, although the primary correlation and network analyses are UK-wide, the LISA and spatially adjusted network analyses are restricted to England. This was a deliberate design choice because those analyses require a consistent OA geography, regional classification and centroid-based neighbour structure. The spatially adjusted network findings should be interpreted as England-specific rather than as UK-wide spatial inference.

Third, harmonisation across the four UK nations necessarily required aggregation of some categories and exclusion of measures that could not be represented consistently, limiting the specificity of some ethnic and contextual classifications. Fourth, the analysis remains subject to the modifiable areal unit problem. Many published ethnic density studies use Lower Super Output Areas or other administratively defined neighbourhood units, whereas this analysis uses harmonised small areas for the UK-wide analyses and Output Areas for the England spatial analyses. The contextual relationships observed here may therefore differ at other spatial scales or under alternative boundary definitions. Finally, the spatially adjusted networks should be interpreted as sensitivity analyses rather than as corrected or superior versions of the original networks. Residualising each variable against English region and its local spatial lag changes the variation being analysed: the resulting networks describe conditional associations within the remaining non-regional and non-locally patterned variation, rather than the full observed neighbourhood context.

# Conclusion

Ethnic density measures should be used with greater conceptual and statistical precision. Researchers should specify what the measure is intended to represent in a given analysis. Across the harmonised UK data, different ethnic density measures were embedded in different configurations of migration, religion, household structure, socioeconomic conditions and spatial settlement, and many of these group-specific associations persisted after spatial adjustment. The implication is not that ethnic density is an invalid exposure, but that its substantive meaning cannot be inferred from the percentage alone. Its use in health research should therefore be tied explicitly to the mechanism under study and interpreted in relation to the wider social and spatial context in which co-ethnic concentration occurs.

More broadly, the paper offers a methodological approach relevant to neighbourhood and urban health research by making contextual co-occurrence itself an object of analysis rather than treating correlated neighbourhood characteristics solely as variables to be adjusted away. At small-area scale, apparently singular contextual measures can be both socially differentiated and strongly spatially patterned. Future studies of neighbourhood effects should therefore examine the composition and spatial organisation surrounding area-level measures before interpreting them as isolated exposures in outcome models. Conditional network and spatial diagnostics offer one approach for identifying the wider contextual characteristics accompanying a neighbourhood measure, and assessing how strongly those relationships are organised geographically.

## Data statement

The UK unified census data product is available to download from the Geographic Data Service at https://data.geods.ac.uk/dataset/unified-uk-census-data. The code pipeline to produce the product is available at https://github.com/GeographicDataService/unified-uk-census-2021-22/.